# Derivative Based Proportionate Approach for Sparse Impulse Response Identification

Murat Babek Salman, and Tolga Ciloglu

*Abstract*— Proportionate type algorithms were developed and excessively used in the echo cancellation problems due to sparse characteristics of the echo channels. In the past, most of the attention was paid to a particular type of proportionate approach, which assigns step-sizes to filter coefficients proportional to the magnitude of the corresponding coefficient. In this letter, we propose a new proportionate type algorithm, which takes dynamic behavior of the estimated filter coefficient into account while assigning individual step-sizes to each coefficient. Proposed algorithm introduces an effective way to assign individual step-sizes using the time derivatives of the filter coefficients. Computational complexity of the proposed algorithm is similar to those of previously proposed algorithms. Simulation results have shown the improvements in the convergence rate achieved by the proposed algorithm.

## I. INTRODUCTION

Adaptive filtering is considered as a feasible solution for system identification problem and it is commonly used in echo cancellation applications in order to model an unknown echo path. In general, echo paths have sparse characteristics such that impulse responses of the echo paths consist of a large number of "minor" coefficients having very small magnitudes compared to a smaller number of "major" coefficients having relatively larger magnitudes. Normalized least mean squares (NLMS) is a well-known method for channel identification problem due to its computational simplicity and robustness. However, for sparse channels NLMS algorithm is an inefficient solution. It yields very slow convergence since sparse channels usually have longer impulse responses compared to regular channels due to long delays in the echo path. To overcome this problem, "proportionate" type algorithms were developed specifically to take the sparse characteristics of the channels into account so that faster convergence can be achieved. In particular, convergences of major coefficients are boosted by applying coefficient specific "proportionate factors" to filter coefficients. The first proportionate type algorithm, proportionate-NLMS (PNLMS) [1], assigns specific step-sizes for filter coefficients roughly proportional to their magnitudes. PNLMS algorithm works well for sparse channels but its performance degrades as the channel impulse response becomes dispersive. Later, a more robust version, improved PNLMS (IPNLMS), was introduced [2]. IPNLMS is a widely used and well-studied algorithm in the context of the echo cancellation problem. It outperforms PNLMS and NLMS for both sparse and dispersive channels.

NLMS type algorithms suffer from performance degradation under colored input and affine projection algorithm (APA) is known to have robustness against colored signals. Therefore, proportionate approach is extended to APA in order to have faster convergence for sparse channels with colored input. PAPA and IPAPA, [3], are the extensions of PNLMS and IPNLMS to APA. Later some other variants of the proportionate type algorithms were proposed such as $\mu$-law IPNLMS, [4], memory improved IPAPA (MIPAPA), [5], and memory improved PAPA with Individual Activation Factors (IAF-MPAPA), [6]-[8].

These conventional proportionate approaches do not take dynamic behavior of the filter coefficients into account while calculating the proportionate factors. However, dynamics of the adaptation provides valuable information about the magnitudes of the variations of the coefficients, which in turn can be used to determine how large step-size for each coefficient is needed at a certain time. Conventional proportionate approaches, on the other hand, control the step-size variation using instantaneous values of the filter coefficients. In [9], gradient controlled IPAPA (GC-IPAPA) was proposed in which time averaged gradient vector is used to calculate proportionate factors. In [10], Coefficient Difference Based IPAPA (D-IPAPA) was proposed in which difference between current filter coefficients and stored filter coefficients is used to calculate proportionate factors.

In this brief, we introduce an algorithm based on the observations of dynamic behavior of the filter coefficients. According to these observations, rapidly changing coefficients correspond to major coefficients; therefore, step-size for such a coefficient is chosen to be a function of the rate of change of that coefficient. Simulation results show that proposed derivative based IPAPA (DB-IPAPA) provides better performance in terms of convergence rate and steady state misalignment compared to MIPAPA, IAF-MPAPA and D-IPAPA.

## II. BACKGROUND

Consider a system identification problem, which includes a channel whose impulse response, $\boldsymbol{h} = [h(0)\ h(1)\ ...\ h(L-1)]^T$ is unknown. Input of the system is denoted by $x(k)$ and $y(k) = \boldsymbol{x}^T(k)\boldsymbol{h} + v(k)$ is the desired signal, where $v(k)$ is the background noise and $\boldsymbol{x}(k) = [x(k)x(k-1)\ ...\ x(k-L+1)]^T$ is the input vector at time $k$. $L$ is the length of the impulse response and superscript $T$ denotes the transpose of a matrix.

APA is a widely used method to identify an unknown system. In [11], the update for the estimate of the filter coefficient vector is given as

$$\boldsymbol{w}(k+1) = \boldsymbol{w}(k) + \mu \boldsymbol{X}(k)(\boldsymbol{X}^T(k)\boldsymbol{X}(k) + \delta \boldsymbol{I})^{-1}\boldsymbol{e}(k), \tag{1}$$

where $\boldsymbol{X}(k) = [\boldsymbol{x}(k)\ \boldsymbol{x}(k-1)\ ...\ \boldsymbol{x}(k-M+1)]$ is the input signal matrix formed by the most recent $M$ input vectors, $M$ is the projection order, $\boldsymbol{w}(k)$ is the estimate of the unknown filter impulse response, $\boldsymbol{e}(k) = \boldsymbol{y}(k) - \boldsymbol{X}^T(k)\boldsymbol{w}(k)$ is the error signal vector, $\boldsymbol{y}(k)$ is the output vector $\boldsymbol{y}(k) = [y(k)\ y(k-1)\ ...\ y(k-M+1)]^T$, $\mu$ is the step-size and $\delta$ is a small positive constant preventing division by zero. It should be noted that for projection order $M = 1$, APA becomes NLMS algorithm.

Proportionate type algorithms, which assign individual step-sizes to filter coefficients, have general form as

$$\boldsymbol{w}(k+1) = \boldsymbol{w}(k) + \mu \boldsymbol{G}(k)\boldsymbol{X}(k)(\boldsymbol{X}^T(k)\boldsymbol{G}(k)\boldsymbol{X}(k) + \delta \boldsymbol{I})^{-1}\boldsymbol{e}(k) \tag{2}$$

where $\boldsymbol{G}(k) = diag(\boldsymbol{g}(k))$ is an $L \times L$ diagonal proportionate matrix, which contains the coefficient specific step-sizes applied to the filter coefficients and $\boldsymbol{g}(k) = [g_0(k), g_1(k), ..., g_{L-1}(k)]^T$ is the proportionate vector. In the literature, different methods were proposed for the calculation of proportionate factors, $g_l(k)$. Conventional IPNLMS (IPAPA) algorithm calculates proportionate factors as

$$g_l(k) = \frac{(1-\alpha)}{2L} + \frac{(1+\alpha)|w_l(k)|}{2\sum_{i=0}^{L-1}|w_i(k)| + \epsilon} \quad \alpha \in [-1, 1], \tag{3}$$

where $\alpha$ is a control variable which adjusts the weights of NLMS (APA) and PNLMS (PAPA) in IPNLMS (IPAPA) and $\epsilon$ is a small positive constant preventing division by zero. Equation (3) assigns smaller step-sizes to the minor coefficients relative to the major coefficients such that deviation of the minor coefficients from their optimal values is avoided, which results in faster convergence. Stemming from [1] and [2], different proportionate type algorithms including, MIPAPA, D-IPAPA, GC-IPAPA, IAF-MPAPA, are developed so that convergence speed of the adaptive filters is improved further.

## III. Proposed Algorithm

In order to derive the proposed algorithm, firstly it should be shown that proportionate factors should be chosen proportional to the difference between the optimal and the current filter coefficients. For this purpose, consider the update of a proportionate type NLMS algorithm

$$\boldsymbol{w}(k+1) = \boldsymbol{w}(k) + \mu \frac{\mathbf{G}(k)\boldsymbol{x}(k)}{\boldsymbol{x}^T(k)\mathbf{G}(k)\boldsymbol{x}(k)} e(k). \tag{4}$$

Equation (4) can be manipulated such that it can be written in terms of coefficient error vector, $\widetilde{\boldsymbol{h}}(k) = \boldsymbol{h} - \boldsymbol{w}(k)$,

$$\widetilde{\boldsymbol{h}}(k+1) = \widetilde{\boldsymbol{h}}(k) - \mu \frac{\mathbf{G}(k)\boldsymbol{x}(k)}{\boldsymbol{x}^T(k)\mathbf{G}(k)\boldsymbol{x}(k)} e(k). \tag{5}$$

Since the normalization term in (5) is common for all filter coefficients, it can be defined as $\sigma_{gx}^2(k) = \boldsymbol{x}^T(k)\mathbf{G}(k)\boldsymbol{x}(k)$. Thus, error for $i^{th}$ coefficient in (5) becomes

$$\widetilde{h}_i(k+1) = \widetilde{h}_i(k) - \mu \frac{g_i(k)x(k-i)}{\sigma_{gx}^2(k)} e(k). \tag{6}$$

In order to prevent the deviation of the filter coefficients from their optimal values, magnitude of the coefficient vector should decrease at each iteration, $|\widetilde{h}_i(k+1)| < |\widetilde{h}_i(k)|$. By replacing $\widetilde{h}_i(k+1)$ with (6), the inequality becomes

$$\left|\widetilde{h}_i(k) - \mu \frac{g_i(k)x(k-i)}{\sigma_{gx}^2} e(k)\right| < |\widetilde{h}_i(k)|. \tag{7}$$

Equation (7) is satisfied only if the following condition is met

$$g_i(k) < \frac{2\sigma_{gx}^2 |\tilde{h}_i(k)|}{\mu |x(k-i)||e(k)|}. \tag{8}$$

Consequently, it is observed that proportionate factor for each coefficient is bounded by the individual coefficient error in order to ensure a monotonic behavior of the coefficient error. Hence, it is concluded that optimal proportionate factors are proportional to the coefficient errors. However, it is impossible to obtain $\tilde{\boldsymbol{h}}(k)$ exactly; hence, a fine approximation of $\tilde{\boldsymbol{h}}(k)$ is required. In order to obtain the approximation, consider several iterations of NLMS algorithm

$$\boldsymbol{w}(k+1) = \boldsymbol{w}(k) + \mu \frac{\boldsymbol{x}(k)}{\boldsymbol{x}^T(k)\boldsymbol{x}(k)} e(k), \tag{9}$$

$$\boldsymbol{w}(k+2) = \boldsymbol{w}(k+1) + \mu \frac{\boldsymbol{x}(k+1)}{\boldsymbol{x}^T(k+1)\boldsymbol{x}(k+1)} e(k+1), \tag{10}$$

$$\vdots$$

$$\boldsymbol{w}(k+N) = \boldsymbol{w}(k+N-1) + \mu \frac{\boldsymbol{x}(k+N-1)}{\boldsymbol{x}^T(k+N-1)\boldsymbol{x}(k+N-1)} e(k+N-1). \tag{11}$$

By summing $\boldsymbol{w}(k+1)$ to $\boldsymbol{w}(k+N)$, one obtains

$$\boldsymbol{w}(k+N) = \boldsymbol{w}(k) + \mu \sum_{j=1}^{N-1} \frac{\boldsymbol{x}(k+j)}{\boldsymbol{x}^T(k+j)\boldsymbol{x}(k+j)} e(k+j). \tag{12}$$

Note that error signal can be written as, $e(k) = \boldsymbol{x}^T(k)\tilde{\boldsymbol{h}}(k) + v(k)$. Then, by inserting $e(n)$ into (12), (12) becomes

$$\boldsymbol{w}(k+N) = \boldsymbol{w}(k)$$
$$+\mu \sum_{j=0}^{N-1} \frac{\boldsymbol{x}(k+j)[\boldsymbol{x}^T(k+j)\tilde{\boldsymbol{h}}(k+j) + v(k+j)]}{\boldsymbol{x}^T(k+j)\boldsymbol{x}(k+j)}. \tag{13}$$

By employing weak law of large numbers, summation in (13) can be approximated by the expectation operator. Consequently, summation in (13) can be approximated as

$$\boldsymbol{w}(k+N) - \boldsymbol{w}(k) \approx \mu N E\left\{\frac{\boldsymbol{x}[\boldsymbol{x}^T \overline{\boldsymbol{h}}(k+N) + v]}{\boldsymbol{x}^T \boldsymbol{x}}\right\}. \tag{14}$$

where $\overline{\boldsymbol{h}}(k+N) \approx \frac{\sum_{j=0}^{N-1} \widetilde{\boldsymbol{h}}(k+j)}{N}$ can be considered as the time averaged coefficient error vector at time $k+N$ assuming that convergence behavior of the filter coefficients does not change significantly within $N$ samples. By using the independent noise assumption, noise related term in (14) becomes zero which reduces the (14) to

$$\boldsymbol{w}(k+N) - \boldsymbol{w}(k) \approx \mu N E\left\{\frac{\boldsymbol{x}\boldsymbol{x}^T \overline{\boldsymbol{h}}(k+N)}{\boldsymbol{x}^T \boldsymbol{x}}\right\}. \tag{15}$$

By utilizing the assumption in [12] and assuming that that input signal, $x(k)$, and coefficient error vector, $\widetilde{\boldsymbol{h}}(k)$, are uncorrelated [13], (15) becomes

$$\boldsymbol{w}(k+N) - \boldsymbol{w}(k) \approx \frac{\mu N E\{\boldsymbol{x}\boldsymbol{x}^T\}\widehat{\boldsymbol{h}}(k+N)}{E\{\boldsymbol{x}^T \boldsymbol{x}\}}, \tag{16}$$

$$\boldsymbol{w}(k+N) - \boldsymbol{w}(k) \approx \frac{\mu N \boldsymbol{R}_x \widehat{\boldsymbol{h}}(k+N)}{N\sigma_x^2}. \tag{17}$$

where $\widehat{\boldsymbol{h}}(k+N) = E\{\overline{\boldsymbol{h}}(k+N)\}$. Since we are dealing with NLMS type algorithms, input signal can be assumed as a white Gaussian signal; hence, $\boldsymbol{R}_x/\sigma_x^2 = \boldsymbol{I}$. Therefore, $\widehat{\boldsymbol{h}}(k)$ can be approximated as,

$$\widehat{\boldsymbol{h}}(k+N) \approx \frac{\boldsymbol{w}(k+N) - \boldsymbol{w}(k)}{\mu}. \tag{18}$$

From (18) it can be concluded that time averaged coefficient error vector is proportional to difference of two instances of the filter coefficients. Therefore, this difference can be employed while calculating proportionate factors. Furthermore, this difference can be inferred as the averaged time derivatives of the filter coefficients since difference in discrete domain corresponds to differentiation in continuous domain.

By utilizing (18), proposed proportionate factors are calculated as follows,

$$g_l(k) = \frac{(1-\alpha)}{2L} + \frac{(1+\alpha)\Delta_l(k)}{K_{mx}(k) + \epsilon}, \quad (19)$$

where $\Delta_l(k)$ is the approximate time derivative of the $l^{th}$ filter coefficient,

$$\Delta_l(k) \triangleq |w_l(k) - w_l(k-N)|, \quad (20)$$

and $K_{mx}(k)$ is the normalization value.

However, it is impractical to use $w_l(k-N)$ directly since it requires storage of additional $N \times L$ filter coefficients. Therefore, in this brief, we propose an alternative method for calculation of $\Delta_l(k)$. $w_l(k-N)$ in (20) is replaced with $\bar{\bar{w}}_l(k)$, which is the stored instance of $w_l(k)$ updated every $L_m$ samples, where $L_m = mL$ and $m$ is a constant. $\bar{\bar{w}}_l(k)$ is defined such that the derivative calculation does not involve the difference of consecutive instances of a particular coefficient.

$$\Delta_l(k) \triangleq |w_l(k) - \bar{\bar{w}}_l(k)|, \quad (21)$$

To get $\bar{\bar{w}}_l(k)$, firstly store $w_l(k)$ periodically at times $k = L_m$, to get $\bar{w}(k)$ as

$$\bar{w}_l(k) \triangleq \begin{cases} w_l(k) & k = nL_m \\ \bar{w}_l(k-1) & k \neq nL_m \end{cases}. \quad n \text{ is an integer} \quad (22)$$

Then $\bar{\bar{w}}(k)$ is obtained by delaying $\bar{w}(k)$ by $L_m$ samples to avoid zero valued derivatives, $\bar{\bar{w}}(k) = \bar{w}(k-L)$, with a less memory requirement, $\bar{\bar{w}}(k)$, can be obtained as

$$\bar{\bar{w}}_l(k) = \begin{cases} \bar{w}_l(k-1) & k = nL_m \\ \bar{\bar{w}}_l(k-1) & k \neq nL_m \end{cases}. \quad n \text{ is an integer} \quad (23).$$

As a result of the development around (22)-(23), separation between coefficients used in the derivative calculation becomes in the range $L_m$ and $2L_m - 1$.

In the proposed algorithm, normalization value, $K_{mx}(k)$, is composed of two parts. The first part is related to the steady-state value of $K_{mx}$ which is the maximum of the absolute values of current filter coefficients. At the steady state, derivative values become negligible compared to the magnitudes of filter coefficients; hence, steady-state values of proportionate factors become very small which helps to avoid random fluctuations of the coefficients as desired. The second part is dominant during the transient period. During the transient period, derivative values may have larger magnitudes compared to filter coefficients especially when impulse response of the unknown system changes during the adaptation. Therefore, it is necessary to control the normalization value in order to avoid impulsive proportionate factors. Hence, normalization value is defined as,

$$W_{mx}[k] \triangleq \max\{|w_0(k)|, |w_1(k)|, \dots, |w_{L-1}(k)|\}, \tag{25}$$

$$\Delta_{mx}[k] \triangleq \max\{\Delta_0(k), \Delta_1(k), \dots, \Delta_{L-1}(k)\}, \tag{26}$$

$$K_{mx}(k) = \frac{\Delta_{mx}^2(k) + W_{mx}^2(k)}{\Delta_{mx}(k) + W_{mx}(k)}. \tag{27}$$

When the adaptation starts, major coefficients move quickly toward their optimum values. Hence, derivative values, $\Delta_l(k)$, of these coefficients become larger than those of minor coefficients. Until major coefficients reach their optimum values, $\Delta_l(k)$ get relatively large values, so major coefficients have larger individual step-sizes. According to (19), as major coefficients get closer to their steady state values, $\Delta_l(k)$'s decrease since $\overline{\overline{w}}_l(k)$ and $w_l(k)$ also get closer values. Therefore, after this time on proportionate vector is dominated by the minor coefficients, whose values are still far from their steady state values. Consequently, convergence of the minor coefficients speeds up.

If computational complexities of DB-IPAPA and other proportionate type algorithms are compared, it can be noticed that DB-IPAPA requires additional memory in order to store filter coefficients. In addition $L$ summations in IPAPA are replaced by $2L$ comparisons in the proposed algorithm. Therefore, it can be stated that proposed algorithm achieves better performance with an acceptable increase in the computational complexity.

IV. SIMULATION RESULTS

Performance of the proposed algorithm is tested and compared with the other algorithms via computer simulations. In particular, the results obtained by the proposed algorithm are compared to those of MIPAPA, [5], D-IPAPA, [9], and IAF-MPAPA, [6] from literature. In all cases, the lengths of the unknown impulse responses are 512. Input signal is either an $AR(1)$ signal, which has a pole at 0.8 or a speech signal. Sample echo path models (EPM) of ITU-T G168 Recommendation [14], padded with zeros, are used as the unknown impulse responses to be identified.

Performances of the algorithms are evaluated by the normalized misalignment (msl) which is defined as, $msl(k) = 20\log_{10}\|\boldsymbol{h} - \boldsymbol{w}(k)\|_2/\|\boldsymbol{h}\|_2 \ dB$. By ensemble averaging over 20 independent realizations of $msl(k)$, misalignment curves are obtained.

9Firstly, performance of the proposed DB-IPAPA is compared to those of the previously proposed MIPAPA, IAF-MPAPA and D-IPAPA for different SNR levels ($30\ dB$ and $15\ dB$) where unknown channel is the first EPM of [14]. Step-size is chosen to be $\mu = 0.15$ for all algorithms with projection order $M = 2$, input signal is an AR(1) signal. In addition, the unknown impulse response is shifted by 50 samples at $15 \times 10^3 th$ iteration in order to examine the tracking performance of the proposed algorithm. Control parameter $\alpha$ is set to 0, the parameter $m$ is set to 1 such that $L_m = L$ for DB-IPAPA, and $P$ is set to $P = 2L$ for D-IPAPA. Initialization vector for IAF-MPAPA is set to $q_m = 10^{-2}/L$. Regularization parameters, $\epsilon$ is set to 0.01 and $\delta = 20\sigma_x^2/2L$ [15], where $\sigma_x^2$ is the input signal power.

Proposed algorithm outperforms previously proposed algorithms as shown in Fig.1 and Fig. 2. It can be observed that superiority of the proposed algorithm is apparent in higher SNR case. DB-IPAPA provides faster convergence speed compared to other algorithms. In addition, the proposed algorithm shows relatively good tracking performance for higher SNR case even if it does not store the locations of the major coefficients unlike conventional proportionate type algorithms. However, the proposed algorithm suffers from slow tracking speed for lower SNR values.

Performance of the proposed algorithm is also tested with speech input signal at sampling rate $8\ kHz$. In this case storing period of the filter coefficients is increased to $L_m = 4L$ samples for DB-IPAPA and $P = 8L$ for D-IPAPA due to high correlation between speech signals and a projection order $M = 8$ is used. The other parameters are kept the same as those in the previous case. Simulation results in Fig. 3 and Fig. 4 show that in the case of speech signal DB-IPAPA algorithm outperforms the other algorithms for both SNR values.

In addition, effects of a relatively dispersive channel on the performance of the algorithms are investigated. In this configuration, the second EPM of [14] is taken as the unknown impulse. For this purpose, AR(1) signal is used as the input signal and SNR is set to $30\ dB$. From Fig. 5, it can be noticed that the proposed algorithm continues to have superior performance in the case of a dispersive echo path.

Another important observation is related to misalignment variance. Since difference values become zero at the steady state, very small step-size is assigned to filter coefficients. Hence, variations of the filter coefficients are minimized in DB-IPAPA, which reduces the misalignment variance at the steady-state which is larger in the classical proportionate type algorithms.

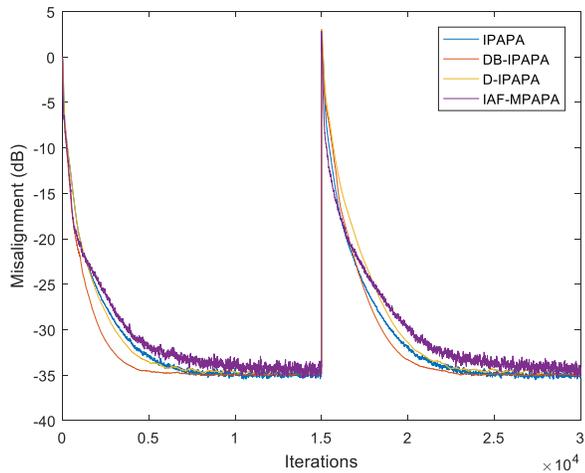

Fig. 1. Misalignment comparisons of IPAPA, DB-IPAPA, D-IPAPA and IAF-MPAPA for EPM-1 and AR(1) input signal with SNR 30 dB.

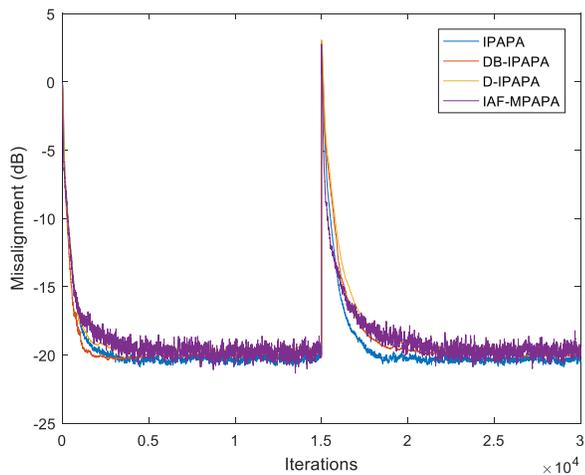

Fig. 2. Misalignment comparisons of IPAPA, DB-IPAPA, D-IPAPA and IAF-MPAPA for EPM-1 and AR(1) input signal with SNR 15 dB.

## V. Conclusions

In this brief, a new approach for the proportionate type adaptive filtering algorithm is proposed. Proposed algorithm is based on the time derivatives of the estimated filter coefficients. Furthermore, a new normalization technique is introduced, which improves convergence speed of the filter coefficients. Simulation results showed that DB-IPAPA outperforms MIPAPA, IAF-MPAPA and D-IPAPA for sparse channels for both AR(1) and speech input signals. In addition, the proposed approach eliminates fluctuations around the optimal point at the steady state.

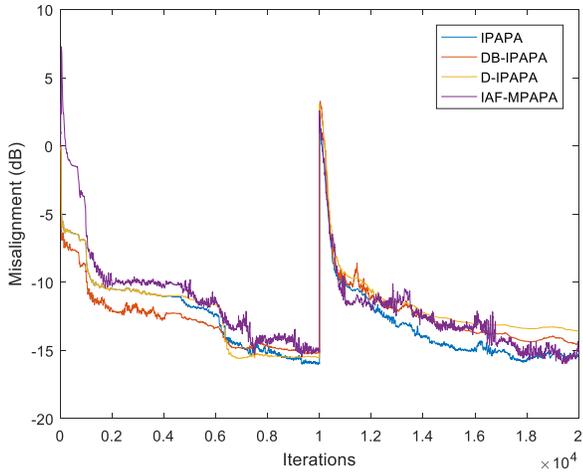

Fig. 3. Misalignment comparisons of IPAPA, DB-IPAPA, D-IPAPA and IAF-MPAPA for EPM-1 and speech input signal with SNR 30 dB.

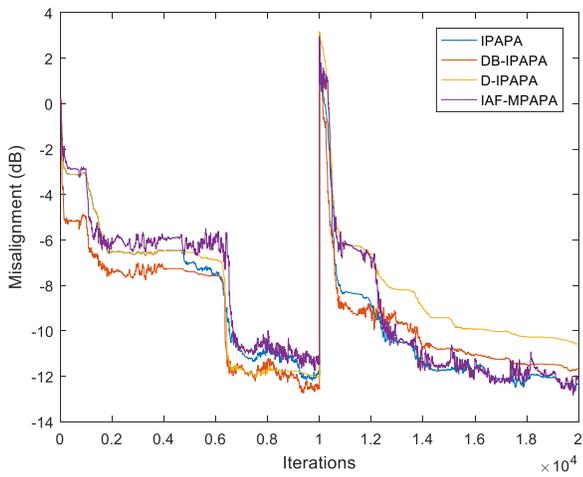

Fig. 4. Misalignment comparisons of IPAPA, DB-IPAPA, D-IPAPA and IAF-MPAPA for EPM-1 and speech input signal with SNR 15 dB.

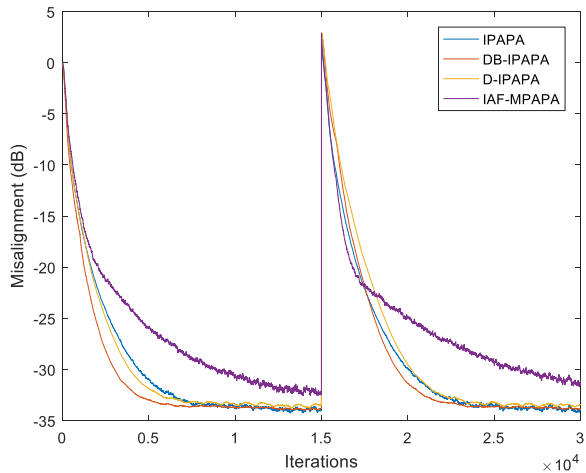

Fig. 5. Misalignment comparisons of IPAPA, DB-IPAPA, D-IPAPA and IAF-MPAPA for EPM-2 and AR(1) input signal with SNR 30 dB

REFERENCES

[1] D. L. Duttweiler, "Proportionate normalized least-mean-squares adaptation in echo cancelers," in *IEEE Transactions on Speech and Audio Processing*, vol. 8, no. 5, pp. 508-518, Sep 2000.

[2] J. Benesty and S. L. Gay, "An improved PNLMS algorithm," in *Proc. IEEE ICASSP*, 2002, pp. II-1881–II-1884.

[3] T. Gansler, J. Benesty, S. L. Gay and M. M. Sondhi, "A robust proportionate affine projection algorithm for network echo cancellation," *2000 IEEE International Conference on Acoustics, Speech, and Signal Processing. Proceedings (Cat. No.00CH37100)*, Istanbul, 2000, pp. II793-II796 vol.2.

[4] Hongyang Deng and M. Doroslovacki, "Improving convergence of the PNLMS algorithm for sparse impulse response identification," in *IEEE Signal Processing Letters*, vol. 12, no. 3, pp. 181-184, March 2005.

[5] C. Paleologu, S. Ciochina and J. Benesty, "An Efficient Proportionate Affine Projection Algorithm for Echo Cancellation," in *IEEE Signal Processing Letters*, vol. 17, no. 2, pp. 165-168, Feb. 2010.

[6] F. d. C. de Souza, O. J. Tobias, R. Seara and D. R. Morgan, "A PNLMS Algorithm With Individual Activation Factors," in *IEEE Transactions on Signal Processing*, vol. 58, no. 4, pp. 2036-2047, April 2010.

[7] H. Zhao, Y. Yu, S. Gao, X. Zeng and Z. He, "Memory Proportionate APA with Individual Activation Factors for Acoustic Echo Cancellation," in *IEEE/ACM Transactions on Audio, Speech, and Language Processing*, vol. 22, no. 6, pp. 1047-1055, June 2014.

[8] Tao Zhang, Hai-Quan Jiao, And Zhi-Chun Lei, "Individual-Activation-Factor Memory Proportionate Affine Projection Algorithm With Evolving Regularization", IEEE Access , March 16, 2017, Digital

Object Identifier 10.1109/ACCESS.2017.2682918.

[9] J. Yang and G. E. Sobelman, "A gradient-controlled proportionate technique for acoustic echo cancellation," *2013 Asilomar Conference on Signals, Systems and Computers*, Pacific Grove, CA, 2013, vol. 2, pp. 1941-1945.

[10] Liu, Ligang. "On Improvement of Proportionate Adaptive Algorithms for Sparse Impulse Response," Ph.D. dissertation, Grad. School of Eng., Kochi Univ. of Tech. Kochi, Japan, (2009).

[11] O. Hoshuyama, R. A. Goubran, and A. Sugiyama, "A generalized proportionate variable step-size algorithm for fast changing acoustic environments," in *Proc. IEEE Int. Conf. Acoust., Speech, Signal Process. (ICASSP)*, Montreal, QC, Canada, 2004, pp. IV–161–IV–164.

[12] Haykin, S. S. (2008). *Adaptive filter theory*. Pearson Education India, pp. 443.

[13] W. Ma, D. Zheng, X. Tong, Z. Zhang and B. Chen, "Proportionate NLMS with Unbiasedness Criterion for Sparse System Identification in the Presence of Input and Output Noises," in *IEEE Transactions on Circuits and Systems II: Express Briefs*.

[14] Digital Network Echo Cancellers 2002, ITU-T Rec. G.168.

[15] C. Paleologu, J. Benesty and F. Albu, "Regularization of the improved proportionate affine projection algorithm," *2012 IEEE International Conference on Acoustics, Speech and Signal Processing (ICASSP)*, Kyoto, 2012, pp. 169-172.